\documentclass[letterpaper]{article} 
\usepackage{aaai25}  
\usepackage{times}  
\usepackage{helvet}  
\usepackage{courier}  
\usepackage[hyphens]{url}  
\usepackage{graphicx} 
\usepackage{natbib}  
\usepackage{caption} 
\usepackage{algorithm}
\usepackage{algorithmic}

\usepackage{newfloat}
\usepackage{listings}
\DeclareCaptionStyle{ruled}{labelfont=normalfont,labelsep=colon,strut=off} 
\floatstyle{ruled}
\newfloat{listing}{tb}{lst}{}
\floatname{listing}{Listing}
\usepackage{comment}
\usepackage{bibentry}
\nobibliography*

\title{Data Ethics in the Fediverse: Analyzing the Role of Instance Policies in Mastodon Research}
\author {
    Mareike Lisker,
    Helena Mihaljevi\'c
}
\affiliations {
    University of Applied Sciences (HTW) Berlin\\
    mareike.lisker@htw-berlin.de, helena.mihaljevic@htw-berlin.de
}
\begin{document}
\maketitle

\begin{abstract}
This article addresses the disconnect between the individual policy documents of Mastodon instances—many of which explicitly prohibit data collection for research purposes—and the actual data handling practices observed in academic research involving Mastodon. We present a systematic analysis of 29 works that used Mastodon as a data source, revealing limited adherence to instance-level policies despite researchers' general awareness of their existence. Our findings underscore the need for broader discussion about ethical obligations in research on alternative, decentralized social media platforms. 
\end{abstract}

\section{Introduction}
The decentralized social media platform Mastodon, launched in 2016, has witnessed a recent surge in scholarly interest. This interest is driven not only by its unique community dynamics, but also by a growing need to explore alternatives to traditional platforms like X, TikTok, and Reddit, whose increasingly restrictive access policies have posed significant barriers to researchers \cite{poudel_navigating_2024,pearson_beyond_2025,corso_what_2024,murtfeldt_rip_2024}.

Mastodon is part of the Fediverse, an open consortium of networks and services that are all built on top of the ActivityPub protocol which facilitates inter-network communication and federation \cite{christine_lemmer-webber_activitypub_2018}.
The Mastodon API is unrestricted in both monetary and technological terms and offers access to over 8.5K federated instances.
While seemingly a solution to data accessibility constraints, the API hides the complexity of the Fediverse in which each instance functions as its own community with distinct rules, privacy policies, and community guidelines. 
These rules delineate the governance and moderation framework of the instance, establish the expected norms for user interactions, outline appropriate posting behaviors, and, in some cases, prescribe the conduct  for other `external actors'  interested in the instance \cite{wahner_dont_2024}.

In 2019, the (non-)compliance with an instance's privacy rules sparked protest within the Mastodon community, leading to the retraction of a paper and the accompanying dataset that contained user-generated data from the platform \cite{zignani_statement_2019}. Several Mastodon admins and users had criticized the publication via an open letter, alleging that it violated the terms of service of at least one instance, as well as the General Data Protection Regulation (GDPR), and that it failed to properly de-identify user data \cite{administrators_scholars_and_users_open_2020}.

Legally, however, an instance's terms and policies are enforceable only between an instance and its registered users, and do not apply to unregistered individuals. Thus, a legally effective violation of the terms of said instance could only have occurred had one of the researchers been a user on that particular instance. 
Nevertheless, ethically, researchers---or as a matter of fact the institutional ethics boards overseeing the research---could still feel obliged to individually review the terms and policies of each instance from which they intend to collect data, ensuring compliance with specific restrictions \cite{roscam_abbing_shifting_2024}. Especially since, as shown by Wähner et al. (\citeyear{wahner_dont_2024}), several instances do in fact include disclaimers against data scraping or academic research in their description.

Mastodon is based on a complex technical infrastructure, fostering a complex social and political environment. 
The mechanism of federation contributes to this, interconnecting the independent instances and allowing users to share and access content across them. Once content is federated to another instance, it is no longer governed by the originating instance's rules and policies.
Furthermore, the official Mastodon API lacks a terms of service or privacy policy, leaving its usage ungoverned. 
Consequently, many researchers may be unaware of the relevance or even existence of individual policies and rules.

In light of this dynamic, our study aims to investigate the following research questions: 
\begin{description}
  \item[\textbf{RQ 1}] How is user-generated data handled in academic research on Mastodon?
  \item[\textbf{RQ 2}] To what extent are the rules and policies of Mastodon instances adhered to in relevant academic research?
\end{description}
To answer these, we conduct a systematic literature review evaluating how Mastodon data as well as instance rules and policies are handled in academic research. Our findings suggest that researchers have limited engagement with policy documents, underscoring the need for increased awareness among researchers, but also other involved parties. 

In the discussion, we propose preliminary recommendations for researchers and ethics committees, as well as for Mastodon software developers, ActivityPub protocol maintainers, and instance administrators, in order to better integrate instance policies into research practice and promote the ethical handling of data from Mastodon and the overall Fediverse.

Although this work critically examines several studies, it is not intended to be a blame-oriented exercise, but rather a constructive contribution to improving social media research practices.

\section{Related Work}

Research involving social media data is shaped by a complex interplay of legal, technical, financial, and ethical constraints that influence every stage of the research life cycle---from data collection to storage, analysis, and sharing. Researchers are typically expected to consider relevant data protection laws, platform terms of service (ToS), and ethical guidelines issued by institutions, funding bodies, or scientific associations that emphasize the importance of respecting user privacy and dignity \cite{franzke_internet_2020,townsend_ethics_2017}. This concern applies even when content is publicly accessible--as users may not expect to become subjects of research or may feel discomfort at being unknowingly included in studies \cite{fiesler_participant_2018}--or when data is anonymized--as effective anonymization is technically and conceptually difficult, making ethical data sharing challenging, if not impossible.

There is an ongoing debate within the academic community about whether it is ethically permissible to violate platform ToS for research purposes \cite{metcalf_where_2016,vitak_ethics_2017,davidson_platform-controlled_2023,fiesler_no_2020}. ToS-based restrictions can hinder the transparency and reproducibility of scientific research or unfairly limit who is able to conduct it---ultimately introducing systemic bias into the scientific process. Moreover, platform ToS are generally designed to protect the interests of platform owners, which may not align with core ethical principles such as justice or beneficence that underpin research ethics \cite{chua_navigating_2022}. As shown by Fiesler et al. \citeyearpar{fiesler_no_2020}, the majority of platforms' ToS documents are formulated in broad and often ambiguous terms, without distinguishing between different contexts and purposes of data collection and processing---considerations that are central to both researchers' ethical considerations \cite{chua_navigating_2022} and users' approval  \cite{gilbert_measuring_2021}.

It is important to note that most discussions around the legality and ethics of ToS violations in the context of publicly available social media data have focused on centralized,  proprietary, commercial platforms such as Twitter/X or Facebook. How these debates translate to decentralized ecosystems like the Fediverse, Mastodon, or individual instances is far from clear. The relationship between users and platform governance differs fundamentally in the Fediverse in comparison to commercial social media networks. Users can choose instances based on their alignment with specific social norms, identities, or political values \cite{colglazier_servers_2024}. As such, the expectations of privacy, consent, and trust are more localized and context-dependent. Violating an instance's rules or collecting data without the community's knowledge may not only undermine user expectations in the instance but also damage trust in the broader ecosystem---an outcome that may carry ethical weight even if the data in question is technically public.
In addition, the Fediverse as a whole emphasizes principles such as autonomy, transparency, and consent; values that resonate with core research ethics frameworks. This raises the question of whether researchers should treat instances more like distinct communities or even research participants, requiring tailored engagement, rather than as abstract data sources.

The practical implications of these tensions became apparent in the previously mentioned retraction of the paper and its accompanying dataset in 2019 \cite{zignani_statement_2019}. The paper faced significant critique, mainly for the inclusion of a verbatim quote that could be traced back to the original post and user, and the violation of the terms of service of a specific Mastodon instance, which in turn conflicted with institutional ethics regulations \cite{administrators_scholars_and_users_open_2020}. 
As shown by Wähner et al. \citeyearpar{wahner_dont_2024}, who conducted an in-depth examination of English rules on Mastodon instances in 2023, 31 out of the 4,371 identified instances explicitly stated in their guidelines that they do not consent to being indexed or researched. 

In response to the retraction case and broader concerns,  Roscam Abbing and Gehl \citeyearpar{roscam_abbing_shifting_2024} released a guide intended to assist researchers transitioning from centralized to decentralized platforms. The guide underscores the instance-specific nature of rules and policies, cautioning researchers that ``The existence of a Mastodon API is not to be confused with the consent of Mastodon users to have their data included in studies.'' \cite[p.2]{roscam_abbing_shifting_2024}
The authors further argue that the Mastodon API artificially ``flattens'' the complexity of the network and ``obscures the role of interfaces altogether'' \cite[p.3]{roscam_abbing_shifting_2024}, offering researchers an overly simplified and potentially misleading view of the platform's structure and norms.

Similar suggestions for more community-centered research ethics have been formulated by scholars examining ethical practices in the context of Reddit—a platform that, like Mastodon, consists of semi-autonomous communities (subreddits) with their own rules, cultures, and moderation practices \cite{fiesler_remember_2024}. A large-scale quantitative study from 2021 analyzed over 700 research papers using Reddit data and found that only 14\% mentioned ethics approval, with many claiming it was unwarranted \cite{proferes_studying_2021}. A follow-up qualitative analysis of the subset that did discuss ethics revealed that considerations were often minimal and framed in procedural terms \cite{fiesler_remember_2024}. Concerningly, nearly 30\% of the studies in \cite{proferes_studying_2021} included direct quotes, about 10\% displayed identifiable usernames, and 7\% explicitly shared datasets. These findings expose a gap between ethical theory and research practice, raising critical questions for how researchers should engage with decentralized platforms. We are not aware of any systematic study that has examined data handling practices specifically in the context of Mastodon. With this work, we aim to contribute to filling that gap.



\section{Data and Methods}
Our focus lies on research efforts that involve the collection of user-generated data from Mastodon, as this is where privacy and ethical concerns become apparent. Specifically, we aimed to examine studies that collected data on toots (posts on Mastodon), user profiles, network links (e.g. follower-followee-relationships), and interaction data (including replies, mentions, or boosts). Note that we excluded (1) papers focusing solely on instance‐level data such as instance policies as, e.g., \cite{wahner_dont_2024,sabo_analysis_2024}, or aggregated usage statistics as e.g., \cite{xavier_evidence-based_2024}, and (2) studies where Mastodon served merely as a recruitment channel for surveys like \cite{lee_uses_2023,gehl_digital_2023}.

\begin{figure}[ht]
\centering 
\includegraphics[width=0.99\linewidth]{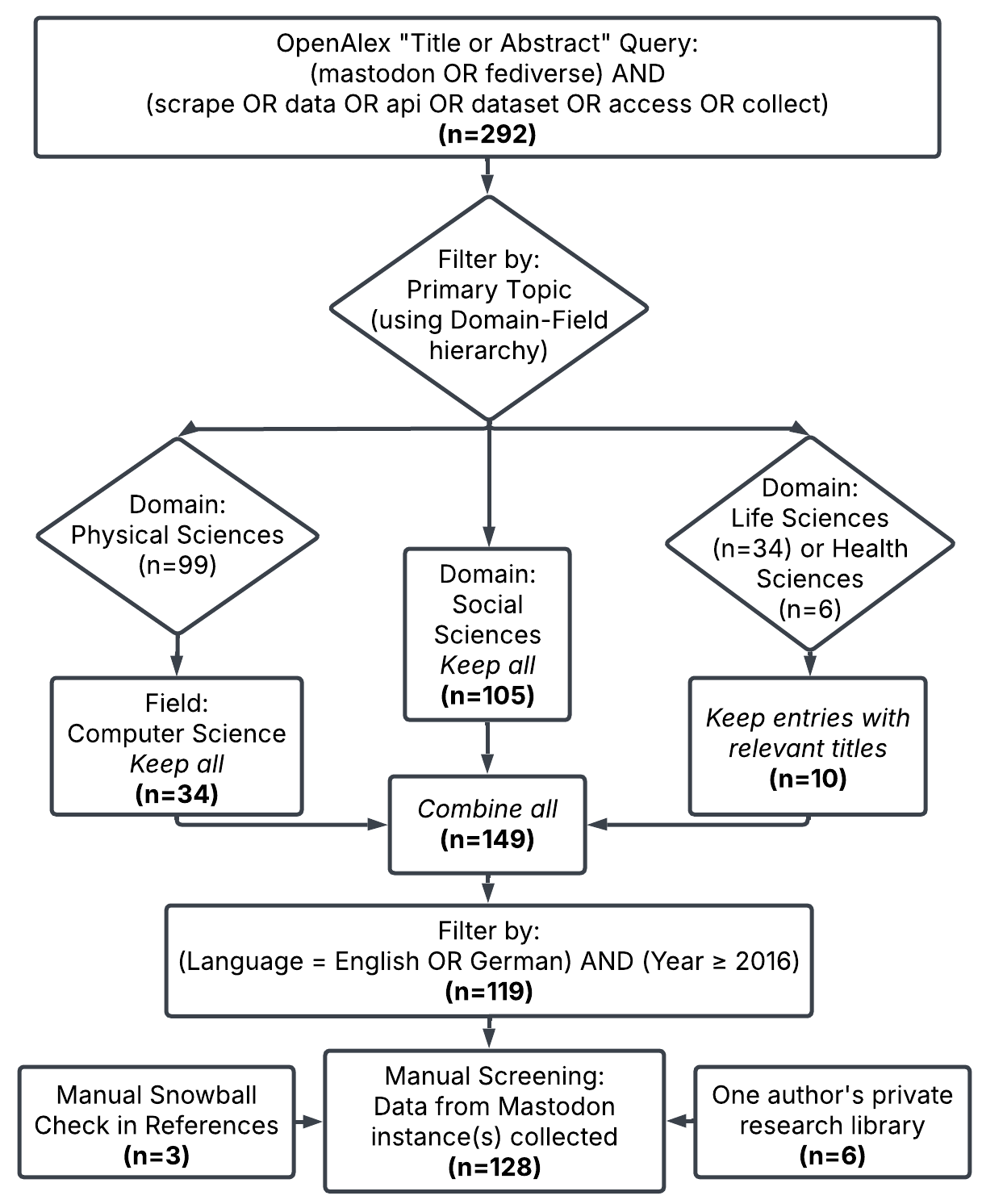}
\caption{Summary of the systematic literature review process}
\label{fig:process}
\end{figure}

In March 2025, we systematically searched the Open Science platform \textit{OpenAlex}, which indexed over 250~million works drawn from Microsoft Academic Graph, Crossref, and other ressources including arXiv and Zenodo \cite{openalexorg_openalex_2025}.  
We queried for titles or abstracts explicitly mentioning \textit{mastodon} or \textit{fediverse} alongside a term indicating data retrieval, resulting in 292 matches. Figure \ref{fig:process} outlines the selection process and query.
Next, we filtered results in OpenAlex using the \textit{primary\_topic} field across  \textit{domain} and \textit{field}.\footnote{OpenAlex tags work with topics using an automated system based on title, abstract, source, and citations. The \textit{primary\_topic} represents the work's highest-scoring topic within a four-level hierarchy.} 
We retained all 105 papers in the `Social Sciences', 34 in `Computer Science' within the `Physical Sciences' (99) after a title-based screening, and added 10 relevant titles from the `Life Sciences' (34) and `Health Sciences' (9) domains after title screening, totaling 149 records.
We then limited the corpus to works in English or German (the authors' working languages) and excluded publications before 2016---reducing the set to 147, and then 119, respectively. A following in-depth content review ensured that each study had collected user-generated data from Mastodon instances as previously specified. This process resulted in 21 relevant publications. 
A manual snowball check of references revealed 3 more papers. Additionally, 6 papers from an author's private library were included, having not been indexed in OpenAlex due to missing abstracts or too recent publication. Finally, our dataset comprises 29 works.

\section{Results}

All 29 included works were published from 2018 onward, with notable growth in 2023 (coinciding with Elon Musk's takeover of Twitter one year prior) and continued expansion in early 2025. Of these, 21 were peer-reviewed, seven were preprints, and one was a dataset.
\textbf{Research objectives} clustered around two main themes: network analysis (8 studies) and inter-platform migration (7), the latter particularly from Twitter to Mastodon. Among others, topics included inter-platform interactions (e.g., Threads), testing follow recommendation algorithms, and dataset compilation for language-specific classifiers. The most common \textbf{data collection method} was the official Mastodon API (15), in some cases requested through tools like \cite{mastodonpy_contributors_mastodonpy_2023}. The service \textit{instances.social} offering an overview over all instances was also used in 8 studies. Other tools included \textit{fediverse.observer} (1) and \textit{mnm.social}\footnote{Note that the domain \textit{mnm.social} has been reassigned as of March 27, 2025.} (1).
\textbf{Dataset size} was reported in 14 papers, though none of them clarified whether only text or also multimedia content was collected. Fifteen papers disclosed how many users they collected data about, either as user profiles, as part of follower/followee-relationships or as part of the collected toots. Only 15 papers specified the \textbf{number of instances} included. Table \ref{tab:instances_toots_users} summarizes the distribution of instances, toots, and users across papers.

\begin{table}[ht]
\centering
\caption{Number of instances, toots, and users by paper count}
\label{tab:instances_toots_users}
\begin{tabular}{lr|lr|lr}
\hline
Instances       & \#  & Toots  & \# & Users & \# \\ \hline
1               & 4   & 600-7.5K & 3 & 1K-10K & 4   \\
10-50           & 3   & 13K-20K  & 2 & 20K-45K & 2 \\
100-400         & 2   & 200K-600K & 2 & 135K-255K & 5 \\
1.4K-8K         & 3   & 1M-6M    & 4 & 480K  & 1 \\
11K-16.3K       & 2   & 67M-104M\footnote{Note that the 104M toots were obtained from Mastodon and the Pleroma network.} & 2 & 1.15M-2M & 3   \\ \hline
\end{tabular}
\end{table}

\textbf{Data spans} ranged from 1 day to 7 years in the 12 papers that specified it, with most covering 1–6 months (8 studies).
\textbf{Instance selection methods} were identified in all studies, with some using multiple criteria. The most common involved filtering by user profiles, hashtags, language, content, or permission via \textit{robots.txt}\footnote{Many instances use a robots.txt-file to instruct web bots on content scraping \cite{the_web_robots_pages_about_2025}, which is however not binding.} (15). Others used snowball sampling (5), scraping from all or any instance(s) (4), top-ranked instances (3), or based selection on user count distribution (1), random choice (1), or a dummy instance (1).

We found that 17 papers addressed \textbf{instance policies} based on a keyword search for \textit{polic-}, \textit{rule}, and \textit{terms}. While several noted the public nature of scraped content, indicating  awareness of the ethical and legal public-private distinction, engagement with policy details varied.
One paper discussed only the platform-wide terms of service; another stressed general adherence to social media ToS. Eleven acknowledged instance-specific governance documents but without clear implications for their methods. Notably, for two of them, the policies were part of the data they collected on moderation and federation dynamics. Also, at least two of the referenced instances prohibit data collection without user consent (as of March 27, 2025); one specifically bans use for AI training.
The authors of the retracted study mentioned before justified their approach by noting that most instances adopt the standard Mastodon privacy policy, which they interpreted as permitting data collection in the absence of an explicit prohibition \cite{zignani_statement_2019}.
A more rigorous approach was the manual review of the ToS for the 125 most frequent instances (covering 95\% of their user sample).


Seven works \textbf{published Mastodon data} at least in part, via GitHub (3), Zenodo (1), Google Drive (1), figshare (1), or an institutional repository (1, link inactive). Three papers promised future publication, but no links were found. Four of the seven published datasets were explicitly \textbf{licensed} under CC-BY-4.0, CC-BY-NC, the GNU General Public License (GPL), and the Mozilla Public License 2.0 (MPL), respectively. Two of those datasets were based on data from single identified instances, but their instance-level rules were not compatible with the applied licenses. The remaining two works selected instances based on hashtags and Twitter profiles, making it likely that the resulting datasets also include content from instances whose policies do not permit redistribution under the stated licenses. One study inaccurately claimed the Mastodon network ``typically follows a Creative Commons license'' \cite{cerisara_multi-task_2018}. 
In one case, cross-referencing toots from the dataset with the specified instance enabled us to de-identify users, rendering anonymization of user IDs ineffective. 

To assess measures regarding user privacy, we searched for the keywords \textit{anonym-} and \textit{pseudo-}, identifying nine papers addressing \textbf{data anonymization} or pseudonymization. Only two of these overlapped with studies that published their data. Three papers claimed to have anonymized data without further elaboration, while six reported anonymizing mainly user IDs. It should, however, be noted that obfuscating or removing user IDs alone is insufficient to prevent potential re-identification when toots are analyzed, as was the case in five of the nine studies.

From 2022 onward, seven studies transferred Mastodon data \textbf{to other APIs}: five used Perspective API, one used DeepAI, two used OpenAI, and another IBM Watson X.

\section{Conclusion and Discussion}

Recent developments highlight the need for greater awareness of the ethical complexities involved in collecting data from decentralized platforms such as Mastodon among researchers. This study analyzed 29 works involving Mastodon user data to better understand the data practices employed. The results obtained are consistent with the findings reported in related work on Reddit  \cite{proferes_studying_2021,fiesler_remember_2024}. While most authors acknowledge instance-level autonomy and the existence of individual policies, few engage with their specific contents and the potential implications for the conducted research. Notably, at least two instances included in the analyzed works in this study explicitly prohibit data collection without consent. Furthermore, seven works have published their data, with four applying inappropriate licenses. A considerable number of works report transferring data to external services like Perspective API or OpenAI, which may retain data for model improvement depending on service utilization.

Our findings suggest that researchers' engagement with instance policies---and by extension, privacy regulations---falls short of addressing the values and norms articulated in those policies. Unlike centralized platforms, instance-specific rules are often more closely aligned with the interests of their user communities. In line with \cite{fiesler_remember_2024} and \cite{roscam_abbing_shifting_2024}, \textbf{we recommend that \underline{researchers} working with Mastodon data familiarize themselves with relevant instance policies and incorporate these into their research design.} While there may be cases where bypassing such rules is ethically defensible, it remains crucial to assess potential harms and benefits for both individuals and communities \cite{fiesler_remember_2024}. For practical guidance, we refer to the aforementioned works.

Particular caution is required when conducting research that actively intervenes in the discourse happening on a social media platform. Recently, an experiment conducted by researchers from the University of Zurich on the ``Change My View'' subreddit on Reddit has led to outrage and sparked substantial discussion about the ethics of social media research \cite{cathleen_ogrady_unethical_2025}. The experiment had been conducted covertly \cite{gehl_ethics_2025}, without informing either the users or the moderators, and without obtaining their consent. The fact that the study was previously approved by the University of Zurich’s Ethics Committee makes it all the more urgent \textbf{for \underline{ethics committees} to continually adapt their audits and guidelines, and be aware of the specifics and complexities of diverse online spaces and stay up to date on social media research ethics.}

At the same time, placing the burden solely on researchers overlooks structural shortcomings. The Mastodon API documentation, for example, includes a section titled ``Playing with public data'', which may suggest that publicly accessible content is freely usable without further ethical scrutiny \cite{mastodon_api_playing_2024}. Similarly, software packages used for data collection vary widely in handling instance-specific rules; some advise consulting individual instance ToS and warn against unauthorized research or third-party data transfers \cite{schoch_rtoot_2024}, while others claim to  address ethical data handling by respecting \textit{robots.txt} and accessing only public instances \cite{zia_mastodoner_2024}, or they do not mention instance-level policies at all \cite{nirmal_sociohub_2023}. 
\textbf{We thus recommend for \underline{developers of software interacting with Mastodon}, e.\,g., the API or wrappers, to extend their documentation by suggesting users to consult the instance guidelines and describing appropriate uses of their tools.}

These inconsistencies indicate a broader need for tools to help developers and platform maintainers more clearly communicate ethical and policy considerations to users, a topic already discussed by Fediverse users in light of an unconsented incorporation of their posts by a new social media network \cite{tilley_maven_2024}.
One way to support more responsible data use is through infrastructural improvements. \textbf{We recommend that \underline{developers maintaining the ActivityPub protocol} implement the technical prerequisites for standardizing and versioning instance rules in a machine-readable format.} This would allow researchers to determine the conditions under which data collection is permitted.

\textbf{Additionally, \underline{instance administrators} should be made aware of the potential for data scraping. They should be encouraged to provide clear, research-relevant, and, preferably, contextualized guidance in their policies.} However, appealing to each administrator only individually poses structural limitations. Therefore, this process should be discussed and implemented collectively among Fediverse maintainers. 

Finally, from the user perspective, enabling user-level permissions would provide the greatest flexibility in expressing individual data-sharing preferences. The Mastodon community has begun discussing proposals inspired by Bluesky's intent signaling in the AT protocol, which could pave the way for more granular consent mechanisms \cite{bluesky-social_0008_2025,hof_fediverse_2025}.

\section{Limitations}
In our exploration of Mastodon-related research, we encountered several limitations that may have affected the comprehensiveness of our findings. First, we observed discrepancies in OpenAlex indexing, including missing abstracts and occasional misclassifications of topic fields or domains. 
Second, relevant studies may have been excluded if they did not contain the selected keywords in their titles or abstracts, or if they employed alternative terminology. Finally, some of our observations are based on keyword searches by skimming the documents, which may have led to the omission of relevant content not captured by the selected terms.

\bibliography{references}

\appendix
\section{Appendix: Surveyed Works}
\begin{itemize}
 \item \bibentry{al-khateeb_dapping_2022}
  \item \bibentry{alvarez_crespo_unveiling_2023}
  \item \bibentry{bittermann_social_2025}
  \item \bibentry{cava_drivers_2023}
  \item \bibentry{cava_network_2022}
  \item \bibentry{cerisara_multi-task_2018}
  \item \bibentry{colglazier_effects_2024}
  \item \bibentry{cursi_herd_2024}
  \item \bibentry{felkner_vulnerability_2024}
  \item \bibentry{gassmann_influence_2024}
  \item \bibentry{jeong_exploring_2024}
  \item \bibentry{jeong_fediversesharing_2025}
  \item \bibentry{jeong_user_2024}
  \item \bibentry{kasnesis_prototype_2020}
  \item \bibentry{la_cava_polarization_2024}
  \item \bibentry{la_cava_understanding_2021}
  \item \bibentry{min_fedilive_2025}
  \item \bibentry{radivojevic_reputation_2024}
  \item \bibentry{raman_challenges_2019}
  \item \bibentry{sabo_analysis_2024}
  \item \bibentry{sassor_dataset_2023}
  \item \bibentry{trienes_recommending_2018}
  \item \bibentry{wang_failed_2024}
  \item \bibentry{zia_collaborative_2025}
  \item \bibentry{zia_flocking_2023}
  \item \bibentry{zignani_follow_2018}
  \item \bibentry{zignani_footprints_2019}
  \item \bibentry{zignani_mastodon_2019}
  \item \bibentry{zulli_rethinking_2020}
\end{itemize}

\end{document}